\documentclass[submission,copyright,creativecommons]{eptcs}
\providecommand{\event}{QAPL 2015} % Name of the event you are submitting to
\usepackage{cite}
\usepackage{pgf,tikz}
\usepackage{pgfplots}
\usetikzlibrary{automata,positioning}
\usepackage[english]{babel}
\usepackage{algorithmic} 
\usepackage{amssymb} 
\usepackage{amsmath}
\usepackage{color} 
\usepackage[T1]{fontenc}
\usepackage{subfig}
\usepackage{graphicx}
\usepackage{paralist}
\usepackage{enumitem}
\usetikzlibrary{shapes.gates.logic.US,trees,positioning,arrows}

\usepackage{tikz-qtree}
\usetikzlibrary{shadows,trees}

\pgfplotscreateplotcyclelist{mycolorlist}{%
blue,every mark/.append style={fill=blue!80!black},mark=*,mark size=1\\%
red,every mark/.append style={fill=red!80!black},mark=square*,mark size=1\\%
brown!60!black,every mark/.append style={fill=brown!80!black},mark=otimes*,mark size=1\\%
black,mark=star,mark size=1\\%
blue,every mark/.append style={fill=blue!80!black},mark=diamond*,mark size=1\\%
red,densely dashed,every mark/.append style={solid,fill=red!80!black},mark=*,mark size=1\\%
brown!60!black,densely dashed,every mark/.append style={
solid,fill=brown!80!black},mark=square*,mark size=1\\%
black,densely dashed,every mark/.append style={solid,fill=gray},mark=otimes*,mark size=1\\%
blue,densely dashed,mark=star,mark size=1,every mark/.append style=solid\\%
red,densely dashed,every mark/.append style={solid,fill=red!80!black},mark=diamond*,mark size=1\\%
}

\title{Time Dependent Analysis with \\Dynamic Counter Measure Trees}
\author{Rajesh Kumar \qquad Dennis Guck\qquad Mari{\"e}lle Stoelinga
\institute{Formal Methods and Tools\\
University of Twente\\
Enschede, Netherlands}
\email{\{r.kumar,d.guck,m.i.a.stoelinga\}@utwente.nl}
}
\def\titlerunning{Time Dependent Analysis with Dynamic Counter Measure Trees}
\def\authorrunning{Kumar, Guck \& Stoelinga}

\begin{document}
\maketitle

\begin{abstract}
The success of a security attack crucially depends on time: the more time available to the attacker, the higher the probability of a successful attack. Formalisms such as Reliability block diagrams, Reliability graphs and Attack Countermeasure trees provide quantitative information about attack scenarios, but they are provably insufficient to model dependent actions which involve costs, skills, and time. In this presentation, we extend the Attack Countermeasure trees with a notion of time; inspired by the fact that there is a strong correlation between the amount of resources in which the attacker invests (in this case time) and probability that an attacker succeeds. This allows for an effective selection of countermeasures and rank them according to their resource consumption in terms of costs/skills of installing them and effectiveness in preventing an attack.
\end{abstract}

\section{Introduction}

Attack trees \cite{Schneier} and its variants\cite{ATformal} are simple yet powerful formalisms to quantitatively express the attack scenarios in a straight forward way. They have been studied extensively in the fields of risk assessment\cite{Paul}, e-voting \cite{2006}, critical infrastructures\cite{Byres13} and socio-technical security\cite{Reddy}. Based on temporal attributes of the leaf nodes, Attack trees can be further classified as static or dynamic, both of which can be further refined to take into only single parameters\cite{ADT}\cite{singleparam} or multi parameters\cite{multiparam}. Static analysis techniques are useful in providing answers such as ``Given an attacker skill set and attempt, what is the probability of attacker to succeed?'' whereas dynamic analysis techniques can answer questions such as ``What is probability that an attacker succeeds with Probability of 0.8 in certain $t$ units?''. They can further be refined by adding defence trees\cite{DT} or countermeasure trees\cite{ACT} where both attacker and defender try to restrict the chances of success of each other.

This presentation involves Dynamic Attack Countermeasure trees (ACTs) which are dynamic attack trees enriched with countermeasures. In this presentation we will focus on the inclusion of countermeasures in dynamic ACTs with AND and OR gates. Further, we evaluate the case study as provided in \cite{ACT} using the ADT toolbox\cite{ADT} as well as ATCalc, an extension of \cite{ATCalc}. As first step we compute the static probabilities by a single parameter bottom up analysis and then extend the model by defining the temporal attributes of the basic attack steps (BAS) to investigate how the attack proceeds over time.

\section{Dynamic Attack Countermeasure Trees}

An Attack Countermeasure tree (ACT) can be seen as a directed acyclic graph. We formally represent an ACT as a tuple $(G,r,L)$ where:
\begin{itemize}
\item $G$ is a directed acyclic graph, i.e $ G=(V,E) $ with $V$ a set of all vertices and $E$ a set of all edges such that $E=\{(v,v') \mid \exists v\in V. v'\in children(v)\}$. 
\item $r\in V$ is the single top root of $(V, E)$. It represents the attacker's goal.
%\item $l:V\rightarrow${\Large$\in$} is a labeling function that assigns to each vertex an AT element i.e $E=(A \cup B) $ where A is a set of logical gates given by { A $  = \forall _{k},A _{k} : A _{k} \in (AND,OR) $ such that $ \forall _{k} \in A , Sons(k) \neq \phi $} and B is a set of basic actions given by B = $ \forall _{k},B _{k}:B _{k} \in {{A _{J}} \parallel {D _{m} \parallel {M _{l}}}} $ such that  $ \forall _{k} \in B,Sons(k) = \phi $ where $ A _{1},A_{2}...,D_{1},D_{2},.... $ and $  M_{1},M_{2}..  $. are basic events $ \in $ B .
\item $L:V\rightarrow\Sigma$ is a labelling function that assigns to each vertex an AT element, i.e $\Sigma=(\text{Gates} \cup \text{Leafs}) $ where $\text{Gates}  = \{AND,OR\}$ is a set of logical gates such that for all $L(v) \in \text{Gates}$, $children(v) \neq~\emptyset $ and  $\text{Leafs} = A _{j} \cup D _{m} \cup M _{k}$ is a set of basic events such that for all $L(v) \in \text{Leafs}$, $children(v) =~\emptyset $ where:
\begin{inparaenum}[(a)]
\item $A_j$ is the set of $j$ attack events;
\item $D_m$ is the set of $m$ detection events;
\item $M_k$ is the set of $k$ mitigation events.
\end{inparaenum}
\end{itemize}
To have a time dependent analysis of ACTs, we need to annotate the leaf nodes with two parameters:
\begin{inparaenum}[(\arabic{enumi})]
\item The probability of a successful execution of an attack step;
\item A random variable \textit{X} that describes the execution of an attack with respect to time.
\end{inparaenum}
Due to the memoryless property of exponential distributions, we define the CDF of the random variable $X$ by:%to be exponential given by :Lambda=-ln(1-p)/t
$$ P[X<t]=1-e^{{-\lambda}t}.$$
The eventual probability of success obtained in step (1) can be used to compute the parameter $\lambda=\frac{-ln(1-p)}{t}$ which defines the timed behaviour of an attacker. Note that since any CDF always approaches $1$ by increasing the time bound $t$, this implies that an attacker always succeeds if he is given a sufficient amount time.

\subsection{Semantics of BAS and Gates}
\begin{description}[style=unboxed,leftmargin=0cm]
\item[Basic Attack Step (BAS):] A BAS is a basic step of an attacker or defender which interacts with a gate. It is activated once it receives an \textit{activation} signal from its parent node. After an exponential delay with rate $\lambda$ the BAS propagates a \textit{success} signal to its parent. Note, that the initial \textit{activation} signal is sent by the top-level node at system start. 

\item[AND:] An AND gate is a conjunction of events. Once it is activated by receiving an activation signal from its parent, it activates its children from left to right. The gate sends out a success signal if all of its attached children are successful. 

\item[OR:] An OR gate is a disjunction of events. The activation is equal to the AND gate. The gate sends out a success signal if any of its attached children is successful.

\item[Countermeasures:] Countermeasures are used to model the defender actions which are used to block associated attack steps. They consist out of two basic events, i.e. \emph{detection} and \emph{mitigation}. The countermeasure is activated on receiving the activation signal from the top node. Once it receives the activation signal it activates only the detection event. After the detection event is successful after an exponential delay by rate $\lambda_1$, the countermeasure gate activates the mitigation event which in turn is successful after an exponential delay by rate $\lambda_2$. A countermeasure gate can be seen from either defender or attacker perspective. From the defender perspective, a countermeasure gate is successful if both detection and mitigation events are successful consecutively. An attacker is interested in an undetected and unmitigated event and his motive is successful if the countermeasure gate fails.
\end{description}

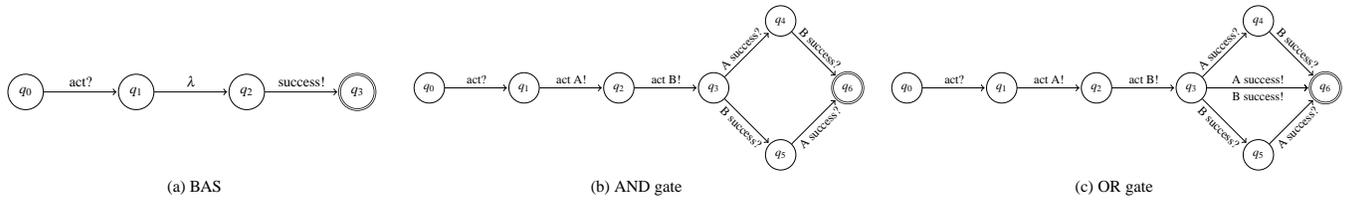
\begin{figure}[t!]
\scalebox{0.70}{
\begin{tabular}{ccc}
\subfloat[BAS]{
\begin{tikzpicture}[shorten >=1pt,node distance=3cm,on grid,auto,scale=0.7, font=\normalsize, every node/.style={transform shape}]
    
     \node[rectangle] (tmp) {};
     \node[state] (q_0) [above of=tmp, node distance=2cm]  {$q_0$}; 
     \node[state] (q_1) [right of=q_0] {$q_1$}; 
     \node[state] (q_2) [right of=q_1] {$q_2$}; 
      \node[state,accepting](q_3) [right of=q_2] {$q_3$};
 
   \path[->]
   
    (q_0) edge  node {act?} (q_1)
    (q_1) edge  node  {$ \lambda $} (q_2)
    (q_2) edge  node  {success!} (q_3);
    
\end{tikzpicture}
\label{fig:bas}
}
&
\subfloat[AND gate]{
\centering
\begin{tikzpicture}[shorten >=1pt,node distance=3cm,on grid,auto,scale=0.6, font=\normalsize, every node/.style={transform shape}]
  
     \node[state] (q_0)  {$q_0$}; 
     \node[state] (q_1) [right of=q_0] {$q_1$}; 
     \node[state] (q_2) [right of=q_1] {$q_2$}; 
     \node[state] (q_3) [right of=q_2] {$q_3$};
      \node[state](q_4) [above right of=q_3] {$q_4$};
  		\node[state](q_5)[below right of=q_3]{$q_5$};
  		\node[state,accepting](q_6)[below right of=q_4]{$q_6$};
  		
   \path[->]
   
    (q_0) edge  node {act?} (q_1)
    (q_1) edge  node  {act A!} (q_2)
    (q_2) edge  node  {act B!} (q_3)
    (q_3) edge  node [sloped, above] {A success?} (q_4)
    (q_3) edge  node [sloped, below]  {B success?} (q_5)
    (q_4) edge  node [sloped, above] {B success?} (q_6)
    (q_5) edge  node  [sloped, below] {A success?} (q_6);  
    
\end{tikzpicture}}
%\hspace{0.5cm} 
&
\subfloat[OR gate]{
\centering
\begin{tikzpicture}[shorten >=1pt,node distance=3cm,on grid,auto,scale=0.6, font=\normalsize, every node/.style={transform shape}]
  
     \node[state] (q_0)  {$q_0$}; 
     \node[state] (q_1) [right of=q_0] {$q_1$}; 
     \node[state] (q_2) [right of=q_1] {$q_2$}; 
     \node[state] (q_3) [right of=q_2] {$q_3$};
      \node[state](q_4) [above right of=q_3] {$q_4$};
  		\node[state](q_5)[below right of=q_3]{$q_5$};
  		\node[state,accepting](q_6)[below right of=q_4]{$q_6$};
  		
   \path[->]
   
    (q_0) edge  node {act?} (q_1)
    (q_1) edge  node  {act A!} (q_2)
    (q_2) edge  node  {act B!} (q_3)
    (q_3) edge  node [above] {A success!} (q_6)
    (q_3) edge  node [below] {B success!} (q_6)
    (q_3) edge  node[sloped, above] {A success?} (q_4)
    (q_3) edge  node [sloped, below]  {B success?} (q_5)
    (q_4) edge  node [sloped, above] {B success?} (q_6)
    (q_5) edge  node  [sloped, below] {A success?} (q_6);  
    
\end{tikzpicture}}
%\hspace{0.5cm}
\\
\end{tabular}
}
\caption{Semantics of BAS and gates in terms of IMCs.}
\end{figure}

\section{Case Study and Interpretation of Results}
\tikzset{font=\scriptsize,
edge from parent fork down,
every node/.style=
    {rectangle,draw,fill=yellow!20,text width=1.0cm,
		text centered,font=\scriptsize,anchor=north}
}

\paragraph{Malicious Insider attack} The ACT for the malicious insider attack (MIA) from \cite{ACT} is depicted in Figure~\ref{fig:act}. The MIA ACT has BAS as well as detection and mitigation events. The countermeasure gates are represented by triangles. We conducted our case study by using the ADT toolbox \cite{ADT} to compare the results to \cite{ACT} and ATCalc \cite{ATCalc} to compute the attack probability over time. %We represent the ACT as a network of BAS and gates which interact with each other. Each BAS and gate is transformed into IMC which perfectly exhibits the timed behaviour. This tree can then transformed into several interacting IMCs which are stepwise composed and minimized \cite{IO} \cite{compos}. This aggregation technique allows for significant state space reduction so that even highly complex models can be handled.

\begin{figure}[ht!]
\scalebox{0.85}{
    \subfloat[Pgoal versus Pleaf]{
         \begin{tikzpicture}[scale=1,font=\tiny,every node/.style={transform shape}]
	\begin{axis}[ymin=0,xmin=0,xmax=1,ymax=1,grid = both,
	xlabel=Pleaf,
	ylabel=Pgoal,
	ylabel shift = 2pt, legend pos=south east,
	cycle list name=mycolorlist]%legend style={at={(0.99,0.11)},anchor=east}]
	
	\addplot table {simulation_case_ADT_1.dat};
	\addlegendentry{No CM}	
	
	\addplot table {simulation_case_ADT_2.dat};
	\addlegendentry{With CM (Detect and Mitig)}	
	
	\addplot table {simulation_case_ADT_3.dat};
	\addlegendentry{Only Detect}	
		\end{axis}
\end{tikzpicture}
        \label{result1}
   }
    \hspace{1cm}
   \subfloat[Pgoal versus Time]{
        \begin{tikzpicture}[scale=1,font=\tiny,every node/.style={transform shape}]
	\begin{axis}[ymin=0,xmin=0,xmax=10,ymax=1,grid = both,
	xlabel=Time,
	ylabel=Pgoal,
	ylabel shift = 1cm, legend pos=south east,
	cycle list name=mycolorlist]%legend style={at={(0.99,0.11)},anchor=east}]
	
	\addplot table {simulation_case_AT_Calc_1.dat};
	\addlegendentry{{Without CM (P=0.05)}}	
	
	\addplot table {simulation_case_AT_Calc_2.dat};
	\addlegendentry{{Without CM (P=0.01)}}	
	
	\addplot table {simulation_case_AT_Calc_3.dat};
	\addlegendentry{{Without CM (P=0.25)}}	
	
	\addplot table {simulation_case_AT_Calc_4.dat};
	\addlegendentry{{CM-Detect (P=0.05)}}	
	
	\addplot table {simulation_case_AT_Calc_5.dat};
	\addlegendentry{{CM-Detect (P=0.1)}}	
	
	\addplot table {simulation_case_AT_Calc_6.dat};
	\addlegendentry{{CM-Detect (P=0.25)}}	
	
	\addplot table {simulation_case_AT_Calc_7.dat};
	\addlegendentry{{CM-Detect+Mitig (P=0.05)}}	
	
	\addplot table {simulation_case_AT_Calc_8.dat};
	\addlegendentry{{CM-Detect+Mitig (P=0.1)}}	
	
	\addplot table {simulation_case_AT_Calc_9.dat};
	\addlegendentry{{CM-Detect+Mitig (P=0.25)}}	
	
	\end{axis}
\end{tikzpicture}
        \label{result3}
    }}
    \caption{$ P_{goal} $ versus (a) $ P_{leaf}  $  (b) Time(sec)}
\end{figure}
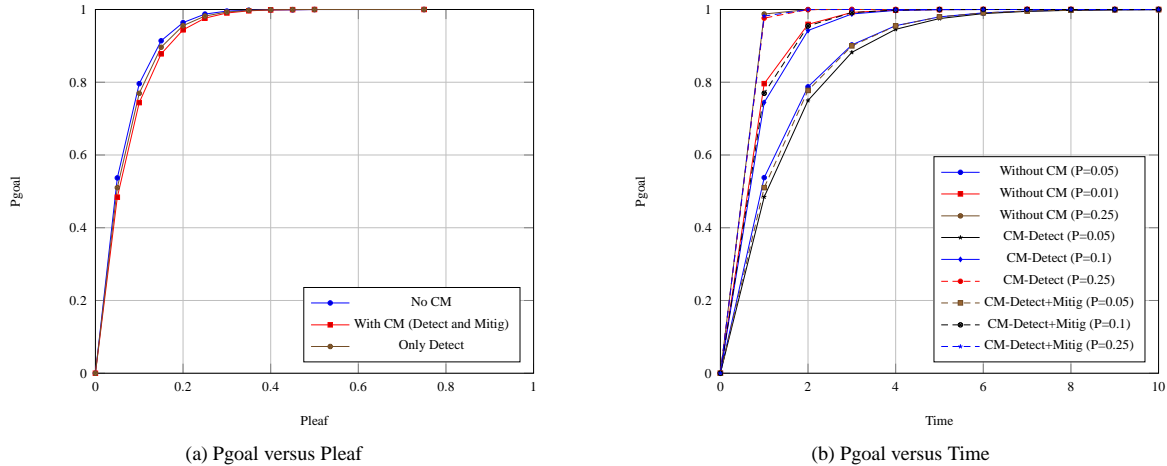

The result in Figure~\ref{result1} is obtained by varying all the probabilities of an attack in the leaf nodes (Pleaf) in the range of $[0,1]$. Figure~\ref{result1} shows that with the countermeasures being in place, the probability of an attack at the root node (Pgoal) first decreases with only detection measures (perfect mitigation) and then increases with detection and mitigation measures in place (imperfect mitigation). Those results are equal to the results obtained in \cite{ACT}. To extend the case study, we consider now the probability of an attack over a time frame of $10$ hours. Figure~\ref{result3} is obtained with different values for Pleaf, fixed at [0.05,0.1,0.25]. The results obtained in Figure~\ref{result3} shows that given an attacker ample time, it is sure that he will eventually be able to reach the goal. Further, it is nicely observable how much more time an attacker needs to reach his attack goal when the detection and mitigation is in place.%This enables us to answer questions like: What is the Probability for attacker to succeed given $'\textit{t}'$ time units?

\begin{figure}[!t]
\begin{center}
\scalebox{0.95}{
\begin{tikzpicture}[level 1/.style={sibling distance=.5cm},
% Gates and symbols style
    and/.style={and gate US, draw,fill=white!60,rotate=90,
		anchor=east,xshift=-1mm},
    or/.style={or gate US,draw,fill=white!60,rotate=90,
		anchor=east,xshift=-1mm},
    be/.style={circle,draw,fill=white!60,anchor=north,
    font=\tiny, text width = 0.8cm, inner sep=0pt},
    tr/.style={buffer gate US,draw,fill=purple!60,rotate=90,
		anchor=east,minimum width=0.8cm},
% Label style
    label distance=3mm,
    every label/.style={blue},
% Event style
    event/.style={rectangle,draw,fill=yellow!20,text width=1.0cm,
		text centered,font=\scriptsize,anchor=north,inner sep=2pt},
% Children and edges style
    edge from parent/.style={draw=black!70},
    edge from parent path={(\tikzparentnode.south) -- ++(0,-1.05cm)
			-| (\tikzchildnode.north)},
    level distance=70pt,
%%  For compatability with PGF CVS add the absolute option:
%   absolute
    scale=0.6, font=\huge,every node/.style={transform shape}]]
\Tree [.\node[event,text width=1.9cm,xshift=2.795cm](G){Malicious Insider Attack Success}; 
[.\node[rectangle,xshift=-3cm](tmp2) {};  
	]         
[.\node[event,text width=1.5cm](A2){Distribution};
    [.\node[event,text width=1.9cm](A21){ File sharing};
    	[.\node[event,text width=1.9cm](A211){Email};
    		[.\node[event,text width=1.9cm](A2111){Local Account};]
    		[.\node[event,text width=1.9cm](A2112){Web based Account};]]
    	[.\node[event,text width=1.9cm](A212){Electronic drop box};
                                 [.\node[event,text width=1.4cm](A2122){Internet};
    									[.\node[event, text width=1.2cm](A21221){Post to news group};]
    									[.\node[event, text width=1.2cm](A2122){Post to website};]]
                                 [.\node[event,text width=1.4cm](A2121){FTP to file server};]]
               [.\node[event,text width=1.9cm](A214){Copy to Media};
                            [.\node[event, text width=1.2cm](A2141){Floppy Disk};]
                            [.\node[event, text width=1.2cm](A2142){CD-Rom};]
                            [.\node[event, text width=1.2cm](A2143){USB-Drive};]]
                [.\node[event,text width=1.5cm](A213){Online Chat};]]]		 
]

\begin{scope}[xshift=13.5cm,yshift=-2.43cm]
\node[rectangle,yshift=.365cm] (tmp3) {};
\Tree 
[.\node[event,text width=1.9cm](A4) {Elevation};  
			[.\node[event,text width=1.9cm](A41){Acquire Admin Previlages};
				   [.\node[event,text width=1.5cm](R){Acquire Password};
		             		  	[.\node[event,text width=1.9cm](A412){Steal Password};
									[.\node[event,text width=1.2cm](A4121){Sniff Network};]
									[.\node[event,text width=1.2cm](A4122){Root Telnet};]]
		                        [.\node[event,text width=1.5cm](S){CM to Steal Password};
									[.\node[event,text width=1.4cm](D412){Track Number of tries at password};]
									[.\node[event,text width=1.4cm](M412){Request admin pin};]]]
		[.\node[event,text width=1.4cm](A413){Sendmail Exploit};]
		[.\node[event,text width=1.4cm](A41){Poor Configuration};]]]	
\end{scope}

\begin{scope}[xshift=-6.3cm,yshift=-2.3cm]
\Tree 
[.\node[event,text width=1.5cm](A1) {Alteration};  
	[.\node[event,text width=1.9cm](P){Manipulation by Virus};
	                   [.\node[event](A12){Launch Virus};]
            		   [.\node[event](Q){Virus CM};
             		  	   [.\node[event,text width=1.5cm](D12){Detect Virus};]
             		  	   [.\node[event,text width=1.5cm](M12){Launch Mitigation};]]]
     [.\node[event, text width=1.9cm](A11){Unauthorized alteration};]]      
\end{scope}

\begin{scope}[xshift=7cm,yshift=-2.43cm]
\node[rectangle,yshift=.365cm] (tmp) {};
\Tree 
[.\node[event,text width=1.9cm](A3){Snooping};
		[.\node[event,text width=1.9cm](A31){Misuse};] 
		[.\node[event,text width=1.5cm](A32){Violation of organizational policy};]]
\end{scope}

\node[left of=tmp, node distance=4.5cm] (tmp4) {};

\path[-] (A3.north) edge[line width=0.2pt,color=black!70] (tmp.center);
\path[-] (A4.north) edge[line width=0.2pt,color=black!70] (tmp3.center);
\path[-] (tmp4.center) edge[line width=0.2pt,color=black!70] (tmp3.center);
		
\node [or]	at (G.south)	[]	{};
\node [or]	 at (A1.south)	[]	{};
\node [and]	at (P.south)	[]	{};
\node [tr]  at (Q.south)	[]	{};
\node [and]  at (A3.south)	[]	{};
\node [or]	at (A21.south)	[]	{};
\node [or]	at (A211.south)	[]	{};
\node [or]	at (A212.south)	[]	{};
\node [or]	at (A2122.south)	[]	{};
\node [or]	at (A214.south)	[]	{};
\node [or]	at (A41.south)	[]	{};
\node [and]	at (R.south)	[]	{};
\node [or]	at (A412.south)	[]	{};
\node [tr]	at (S.south)	[]	{};

\end{tikzpicture}}
\end{center}
\caption{ACT of the malicious insider attack.}
\label{fig:act}
\end{figure}
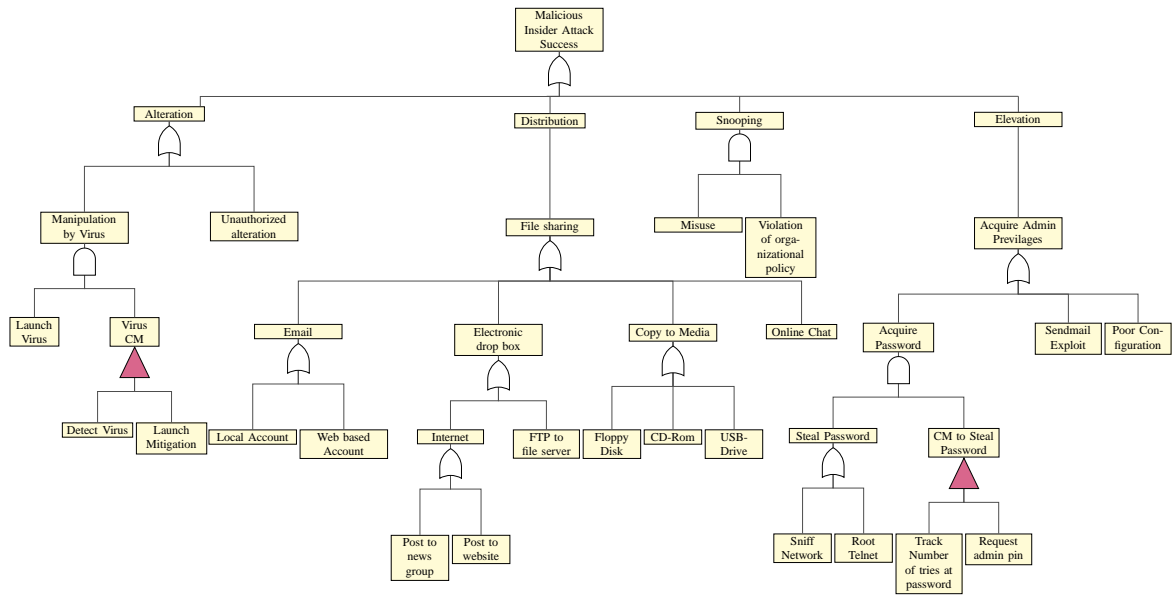

\section{Conclusion}
We presented the inclusion of time in Attack Countermeasure trees and provided a case study to show the applicability of this approach. This enables us to answer questions like: What is the probability for an attacker to succeed given $'\textit{t}'$ time units by integrating new countermeasures? In future work we consider shared attack scenarios as well as extend the dynamic ACTs with:
\begin{inparaenum}[(\arabic{enumi})]
\item Sequential AND and Sequential OR gates which can model the casual dependencies of attack steps, i.e. an attack can only take place if attack steps are executed in a certain order;
\item  Probabilistic gates which activates the BAS and countermeasure events with discrete probabilities, i.e. attacks as well as countermeasures are only executed with a certain probability.%This will enable to model attacks with attack steps which are well studied in literature and can be assigned frequency.
\end{inparaenum}

\paragraph{Acknowledgement.} This work has been supported by the EU FP7 project TREsPASS (318003) and by the STW-ProRail partnership program ExploRail under the project ArRangeer (12238). %We acknowledge our cooperation with Movares in the ArRangeer project.

\bibliographystyle{eptcs}
\bibliography{mybiblo}

\begin{thebibliography}{10}
\providecommand{\bibitemdeclare}[2]{}
\providecommand{\surnamestart}{}
\providecommand{\surnameend}{}
\providecommand{\urlprefix}{Available at }
\providecommand{\url}[1]{\texttt{#1}}
\providecommand{\href}[2]{\texttt{#2}}
\providecommand{\urlalt}[2]{\href{#1}{#2}}
\providecommand{\doi}[1]{doi:\urlalt{http://dx.doi.org/#1}{#1}}
\providecommand{\bibinfo}[2]{#2}

\bibitemdeclare{inproceedings}{ATCalc}
\bibitem{ATCalc}
\bibinfo{author}{F.~\surnamestart Arnold\surnameend}, \bibinfo{author}{A.F.E.
  \surnamestart Belinfante\surnameend}, \bibinfo{author}{F.I. \surnamestart
  van~der Berg\surnameend}, \bibinfo{author}{D.~\surnamestart Guck\surnameend}
  \& \bibinfo{author}{M.I.A. \surnamestart Stoelinga\surnameend}
  (\bibinfo{year}{2013}): \emph{\bibinfo{title}{DFTCalc: {A} Tool for Efficient
  Fault Tree Analysis}}.
\newblock In: {\sl \bibinfo{booktitle}{Proc. of the 32nd Int. Conf. on Computer
  Safety, Reliability, and Security {SAFECOMP}}}, pp.
  \bibinfo{pages}{293--301}, \doi{10.1007/978-3-642-40793-2\_27}.

\bibitemdeclare{article}{DT}
\bibitem{DT}
\bibinfo{author}{S.~\surnamestart Bistarelli\surnameend},
  \bibinfo{author}{F.~\surnamestart Fioravanti\surnameend},
  \bibinfo{author}{P.~\surnamestart Peretti\surnameend} \&
  \bibinfo{author}{F.~\surnamestart Santini\surnameend} (\bibinfo{year}{2012}):
  \emph{\bibinfo{title}{Evaluation of complex security scenarios using defense
  trees and economic indexes}}.
\newblock {\sl \bibinfo{journal}{J. Exp. Theor. Artif. Intell.}}
  \bibinfo{volume}{24}(\bibinfo{number}{2}), pp. \bibinfo{pages}{161--192},
  \doi{10.1080/13623079.2011.587206}.

\bibitemdeclare{article}{Byres13}
\bibitem{Byres13}
\bibinfo{author}{E.~\surnamestart Byres\surnameend} (\bibinfo{year}{2013}):
  \emph{\bibinfo{title}{The air gap: {SCADA}'s enduring security myth}}.
\newblock {\sl \bibinfo{journal}{Commun. {ACM}}}
  \bibinfo{volume}{56}(\bibinfo{number}{8}), pp. \bibinfo{pages}{29--31},
  \doi{10.1145/2492007.2492018}.

\bibitemdeclare{article}{singleparam}
\bibitem{singleparam}
\bibinfo{author}{B.~\surnamestart Kordy\surnameend},
  \bibinfo{author}{S.~\surnamestart Mauw\surnameend} \&
  \bibinfo{author}{P.~\surnamestart Schweitzer\surnameend}
  (\bibinfo{year}{2012}): \emph{\bibinfo{title}{Quantitative Questions on
  Attack-Defense Trees}}.
\newblock {\sl \bibinfo{journal}{CoRR}} \bibinfo{volume}{abs/1210.8092}.
\newblock \urlprefix\url{http://arxiv.org/abs/1210.8092}.

\bibitemdeclare{article}{ATformal}
\bibitem{ATformal}
\bibinfo{author}{B.~\surnamestart Kordy\surnameend},
  \bibinfo{author}{L.~\surnamestart Pietre{-}Cambacedes\surnameend} \&
  \bibinfo{author}{P.~\surnamestart Schweitzer\surnameend}
  (\bibinfo{year}{2013}): \emph{\bibinfo{title}{{DAG}-Based Attack and Defense
  Modeling: Don't Miss the Forest for the Attack Trees}}.
\newblock {\sl \bibinfo{journal}{CoRR}} \bibinfo{volume}{abs/1303.7397}.
\newblock \urlprefix\url{http://arxiv.org/abs/1303.7397}.

\bibitemdeclare{inproceedings}{ADT}
\bibitem{ADT}
\bibinfo{author}{Barbara \surnamestart Kordy\surnameend},
  \bibinfo{author}{Piotr \surnamestart Kordy\surnameend},
  \bibinfo{author}{Sjouke \surnamestart Mauw\surnameend} \&
  \bibinfo{author}{Patrick \surnamestart Schweitzer\surnameend}
  (\bibinfo{year}{2013}): \emph{\bibinfo{title}{ADTool: Security Analysis with
  Attack-Defense Trees}}.
\newblock In: {\sl \bibinfo{booktitle}{Quantitative Evaluation of Systems -
  10th International Conference, {QEST} 2013, Buenos Aires, Argentina, August
  27-30, 2013. Proceedings}}, pp. \bibinfo{pages}{173--176},
  \doi{10.1007/978-3-642-40196-1\_15}.

\bibitemdeclare{inproceedings}{2006}
\bibitem{2006}
\bibinfo{author}{E.~\surnamestart Lazarus\surnameend}, \bibinfo{author}{D.L.
  \surnamestart Dill\surnameend} \& \bibinfo{author}{J.~\surnamestart
  Epstein\surnameend} (\bibinfo{year}{2011}): \emph{\bibinfo{title}{Applying a
  Reusable Election Threat Model at the County Level}}.
\newblock In: {\sl \bibinfo{booktitle}{Electronic Voting Technology Workshop /
  Workshop on Trustworthy Elections{EVT/WOTE}}}, pp. \bibinfo{pages}{12--12}.
\newblock \urlprefix\url{http://dl.acm.org/citation.cfm?id=2028012.2028024}.

\bibitemdeclare{inproceedings}{multiparam}
\bibitem{multiparam}
\bibinfo{author}{A.~\surnamestart Lenin\surnameend} \&
  \bibinfo{author}{A.~\surnamestart Buldas\surnameend} (\bibinfo{year}{2014}):
  \emph{\bibinfo{title}{Limiting Adversarial Budget in Quantitative Security
  Assessment}}.
\newblock In: {\sl \bibinfo{booktitle}{Proc. of the 5th Int. Conf. on Decision
  and Game Theory for Security (GameSec)}}, pp. \bibinfo{pages}{155--174},
  \doi{10.1007/978-3-319-12601-2\_9}.

\bibitemdeclare{article}{Paul}
\bibitem{Paul}
\bibinfo{author}{S.~\surnamestart Paul\surnameend} \& \bibinfo{author}{R.l
  \surnamestart Vignon{-}Davillier\surnameend} (\bibinfo{year}{2014}):
  \emph{\bibinfo{title}{Unifying traditional risk assessment approaches with
  attack trees}}.
\newblock {\sl \bibinfo{journal}{J. Inf. Sec. Appl.}}
  \bibinfo{volume}{19}(\bibinfo{number}{3}), pp. \bibinfo{pages}{165--181},
  \doi{10.1016/j.jisa.2014.03.006}.

\bibitemdeclare{inproceedings}{Reddy}
\bibitem{Reddy}
\bibinfo{author}{K.~\surnamestart Reddy\surnameend}, \bibinfo{author}{H.S.
  \surnamestart Venter\surnameend}, \bibinfo{author}{M.S. \surnamestart
  Olivier\surnameend} \& \bibinfo{author}{I.~\surnamestart Currie\surnameend}
  (\bibinfo{year}{2008}): \emph{\bibinfo{title}{Towards Privacy Taxonomy-Based
  Attack Tree Analysis for the Protection of Consumer Information Privacy}}.
\newblock In: {\sl \bibinfo{booktitle}{Proc of the 6th Annual Conf. on Privacy,
  Security and Trust ({PST})}}, pp. \bibinfo{pages}{56--64},
  \doi{10.1109/PST.2008.18}.

\bibitemdeclare{article}{ACT}
\bibitem{ACT}
\bibinfo{author}{Arpan \surnamestart Roy\surnameend},
  \bibinfo{author}{Dong~Seong \surnamestart Kim\surnameend} \&
  \bibinfo{author}{Kishor~S. \surnamestart Trivedi\surnameend}
  (\bibinfo{year}{2012}): \emph{\bibinfo{title}{Attack countermeasure trees
  {(ACT):} towards unifying the constructs of attack and defense trees}}.
\newblock {\sl \bibinfo{journal}{Security and Communication Networks}}, pp.
  \bibinfo{pages}{929--943}, \doi{10.1002/sec.299}.
\newblock \urlprefix\url{http://dx.doi.org/10.1002/sec.299}.

\bibitemdeclare{book}{Schneier}
\bibitem{Schneier}
\bibinfo{author}{B.~\surnamestart Schneier\surnameend} (\bibinfo{year}{2008}):
  \emph{\bibinfo{title}{Schneier on security}}.
\newblock \bibinfo{publisher}{Wiley}.

\end{thebibliography}

\end{document}